\ifx\iflatexml\undefined
  \RequirePackage{snapshot} %
\fi
\documentclass[conference]{IEEEtran}

\usepackage{microtype}

\usepackage[utf8]{inputenc}

\usepackage[T1]{fontenc}
\usepackage{amsmath}
\usepackage{amssymb}
\usepackage{bm}
\usepackage{xspace}
\usepackage{blindtext}
\usepackage[english]{babel}
\usepackage{booktabs}
\usepackage{multirow}

\usepackage{hyperref}

\usepackage{todonotes}

\ifx\iflatexml\undefined
  \usepackage{siunitx}
\else
  \usepackage{siunitx}[=v2]

  \newcommand\qty[2]{\SI[per=slash]{#1}{#2}}
  \newcommand\unit[1]{\si{#1}}
  \newcommand\qtyrange[3]{\qty{#1}{#3} to \qty{#2}{#3}}
\fi

\DeclareSIUnit\ssjops{\text{ssj\_ops}}

\usepackage{subfig}

\ifx\iflatexml\undefined
  \usepackage{dblfloatfix}
\fi

\usepackage{csquotes}

\usepackage[inline]{enumitem}

\usepackage{flushend}

\newenvironment{enumerateinline}{\begin{enumerate*}[label=(\alph*)]}{\end{enumerate*}}

\makeatletter
\newcommand\thefontsize[1]{{#1 The current font size is: \f@size pt\par}}
\makeatother

\newcommand\extrapolatedidlequotient{extrapolated idle quotient\xspace}
\newcommand\eiq{\extrapolatedidlequotient}

\newcommand{\parsedRejectedBySPEC}{40}
\newcommand{\parsedCoreThreadsInconsistent}{5}
\newcommand{\parsedCoreThreadsImplausible}{1}
\newcommand{\parsedDateAmbiguous}{3}
\newcommand{\parsedDateImplausible}{4}
\newcommand{\parsedCPUAmbiguous}{3}
\newcommand{\parsedNodeCountMissing}{1}
\newcommand{\parsedTotalExamined}{1017}

\newcommand{\parsedTotalAccepted}{960}
\newcommand{\unfilteredYearFirst}{2005}
\newcommand{\unfilteredYearDipFirst}{2013}

\newcommand{\unfilteredYearDipLast}{2017}
\newcommand{\unfilteredYearDipAfterLast}{2018}
\newcommand{\unfilteredYearLast}{2023}

\newcommand{\unfilteredRunsPerYearDipMean}{15.2}

\newcommand{\unfilteredRunsPerYearAll}{44.2}
\newcommand{\unfilteredLinuxPercentageUpToDipEnd}{2.2}
\newcommand{\unfilteredLinuxPercentageAfterDipEnd}{36.3}
\newcommand{\unfilteredAMDPercentageUpToDipEnd}{13.0}
\newcommand{\unfilteredAMDPercentageAfterDipEnd}{31.3}
\newcommand{\unfilteredRunsNeitherAMDNorIntel}{9}
\newcommand{\unfilteredRunsNotSever}{6}
\newcommand{\unfilteredRunsOnlyCommonCfgs}{269}
\newcommand{\unfilteredRunsAfterFiltering}{676}

\newcommand{\unfilteredUpToDipLastWindowsPercentageMinimum}{97}

\newcommand{\powerRecentYearsFirst}{2022}

\newcommand{\powerOldYearsLast}{2010}
\newcommand{\powerRecentYearsMeanFullLoadPower}{303.3}
\newcommand{\powerOldYearsMeanFullLoadPower}{119.0}
\newcommand{\powerFullLoadFactor}{2.5}

\newcommand{\powerTwentyPercentLoadFactor}{1.8}

\newcommand{\powerSeventyPercentLoadFactor}{2.2}
\newcommand{\effRunsByAMDInTopHundred}{98}
\newcommand{\idleFirstYear}{2006}
\newcommand{\idleFirstYearMeanIdleFractionPercent}{70.1}
\newcommand{\idleMinimumIdleFractionYearlyPercent}{15.7}
\newcommand{\idleMinimumIdleFractionYearlyYear}{2017}
\newcommand{\idleMaximumIdleFractionYearlyRegressionYear}{2024}
\newcommand{\idleMaximumIdleFractionYearlyRegressionPercent}{25.7}
\newcommand{\lineupsCutoffStartYear}{2021}
\newcommand{\lineupsMeanCoresAMD}{85.8}
\newcommand{\lineupsMeanCoresIntel}{39.5}
\newcommand{\lineupsFreqGHzMeanAMD}{2.3}

\newcommand{\lineupsFreqGHzSDAMD}{0.3}
\newcommand{\lineupsFreqGHzSDIntel}{0.5}

\makeatletter
\def\mycopyrighttext{
  \textcopyright 2024 IEEE.
  Personal use of this material is permitted.
  Permission from IEEE must be obtained for all other uses, in any current or future media, including reprinting/republishing this material for advertising or promotional purposes, creating new collective works, for resale or redistribution to servers or lists, or reuse of any copyrighted component of this work in other works.
  The definitive version of record of this paper was published as \href{https://doi.org/10.1109/CLUSTERWorkshops61563.2024.00020}{DOI 10.1109/CLUSTERWorkshops61563.2024.00020}.
}
\def\ps@IEEEtitlepagestyle{%
  \def\@oddfoot{\mycopyrightnotice}%
  \def\@evenfoot{}%
}
\def\mycopyrightnotice{%
  \begin{minipage}{\textwidth}
  \scriptsize
  \mycopyrighttext
  \end{minipage}
}
\makeatother

\begin{document}

\title{16 Years of SPEC Power: An~Analysis~of~x86~Energy~Efficiency~Trends}

\author{
    \IEEEauthorblockN{Hannes Tröpgen, Robert Schöne, Thomas Ilsche, Daniel Hackenberg}
    \IEEEauthorblockA{
        ZIH, CIDS, TU Dresden, 01062 Dresden, Germany \\
        \{hannes.troepgen, robert.schoene, thomas.ilsche, daniel.hackenberg\}@tu-dresden.de
    }
}

\ifx\DraftModeOn\undefined

\newcommand{\todoti}[1]{}
\newcommand{\todomb}[1]{}
\newcommand{\todors}[1]{}
\newcommand{\todoag}[1]{}
\newcommand{\tododh}[1]{}
\newcommand{\todoht}[1]{}

\newcommand{\figref}[1]{Figure~\ref{fig:#1}}
\newcommand{\tabref}[1]{Table~\ref{tab:#1}}
\newcommand{\secref}[1]{Section~\ref{sec:#1}}
\newcommand{\lstref}[1]{Algorithm~\ref{alg:#1}}

\newcommand{\papertarget}[2]{}

\renewcommand{\todo}[1]{}
\newcommand{\todoblock}[1]{}

\else

\newcommand{\todoti}[1]{\todo[color=yellow!60,inline,size=\small]{Thomas: #1}}
\newcommand{\todomb}[1]{\todo[color=cyan!60,inline,size=\small]{Mario: #1}}
\newcommand{\todors}[1]{\todo[color=green!60,inline,size=\small]{Robert: #1}}
\newcommand{\tododh}[1]{\todo[color=red!60,inline,size=\small]{Daniel: #1}}
\newcommand{\todoht}[1]{\todo[color=pink!60,inline,size=\small]{Hannes: #1}}

\newcommand{\figref}[1]{\textcolor{red}{Figure~\ref{fig:#1}}}
\newcommand{\tabref}[1]{\textcolor{red}{Table~\ref{tab:#1}}}
\newcommand{\secref}[1]{\textcolor{red}{Section~\ref{sec:#1}}}
\newcommand{\lstref}[1]{\textcolor{red}{Algorithm~\ref{alg:#1}}}

\newcommand{\todoblock}[1]{\todo[inline]{#1}}

\fi

\newcommand{\specpower}[0]{SPEC Power\xspace}
\newcommand{\specpowernext}[0]{SPECpowerNext\xspace}
\newcommand{\speccpu}[0]{SPEC CPU\xspace}
\newcommand{\specomp}[0]{SPEC OMP 2012\xspace}
\newcommand{\specaccel}[0]{SPEC ACCEL\xspace}

\maketitle

\begin{abstract}

The \specpower benchmark offers valuable insights into the energy efficiency of server systems, allowing comparisons across various hardware and software configurations.
Benchmark results are publicly available for hundreds of systems from different vendors, published since 2007.
We leverage this data to perform an analysis of trends in x86 server systems, focusing on power consumption, energy efficiency, energy proportionality and idle power consumption.
Through this analysis, we aim to provide a clearer understanding of how server energy efficiency has evolved and the factors influencing these changes.

\end{abstract}


\ifx\iflatexml\undefined
\else
\mycopyrighttext
\fi

\section{Introduction}
\label{sec:intro}

SPECpower\_ssj~2008\footnote{\emph{SPECpower\_ssj~2008} is the first and so far only release of the \emph{SPEC Power benchmark suite} released by the \emph{Standard Performance Evaluation Corporation} (SPEC).} is the most prominent server energy efficiency benchmark.
Its rigorous methodology and healthy benchmark submission review process have led to 16 years of continuous benchmark submissions and corresponding published data.
These results allow hardware vendors to rank and promote their systems with respect to energy efficiency, measured in \unit{\ssjops\per\watt}.
This metric gives customers an idea of how much computing they get for each invested Joule of energy, where a lower power consumption can increase it as well as a higher processing performance.
\figref{feature_share_over_years_unfiltered} illustrates some strengths of the benchmark:
Due to the simplicity and scalability of the benchmark, server systems with multiple sockets and/or nodes can be measured. %
The workload can also be executed on different operating systems (OS) and different hardware, even though non-x86 processors are rare, and up to \unfilteredYearDipLast{}, more than \qty{\unfilteredUpToDipLastWindowsPercentageMinimum}{\percent} of results use Windows.

Based on the benchmark results available via the SPEC~Power website, we track the performance and power efficiency of x86 processors over the previous 16 years.
We analyze data for different load levels to evaluate energy proportionality, as well as active idle power consumption trends.
	
\begin{figure}[bp!]
  \centering
  \vspace{-1em}
  \includegraphics[width=\columnwidth]{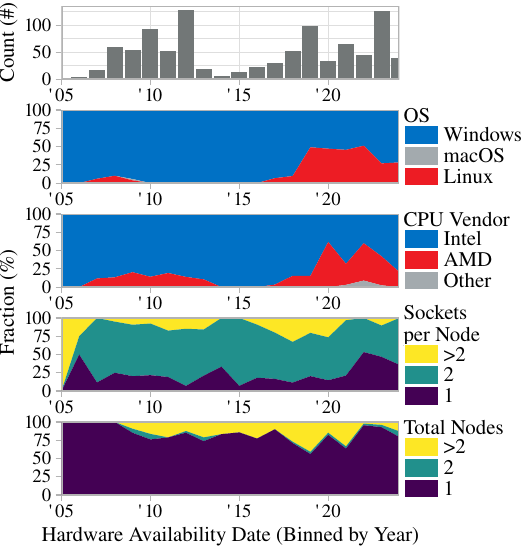}
  \caption{Share of features on all \num{\parsedTotalAccepted} successfully parsed (unfiltered) SPECpower\_ssj~2008 results (as of June 2024)}
  \label{fig:feature_share_over_years_unfiltered}
\end{figure}

\section{Background and Related Work}
\label{sec:background}
\label{sec:related}
SPECpower\_ssj~2008~\cite{specpower_ssjdesign,specpower_andtools} is designed to measure \textquote{the performance and power consumption of servers}.
It consists of an integer-heavy transactional Java-based client/server workload with six differently weighted transaction types.
A calibration phase is used to determine the maximum throughput of the system under test (SUT), which runs the server side.
Partial loads of $\qty{10}{\percent}, \qty{20}{\percent}, \ldots, \qty{90}{\percent}$ are created by scaling down the number of transactions proportionally.
This allows the SUT to apply power-saving mechanisms~\cite[Section Power-Saving Techniques]{Gough2015} and can be used to analyze energy proportionality~\cite{specpower_proportionality,specpower_proportionality2}.
The test regime also includes a 0\% load point, which greatly helps to record, track, and optimize \emph{active idle} power consumption.  %

Based on the SPEC Power methodology, vendors and performance engineers generated hundreds of reports with \num{\parsedTotalExamined} being published on the SPEC website\footnote{\url{https://www.spec.org/power_ssj2008/results/}} at the time of writing.
This long history of vendor-submitted data distinguishes it from other energy efficiency benchmarks, e.g., Green500~\cite{green500} or \specomp~\cite{SPEComp2012}.
The SPECpower committee\footnote{\url{https://www.spec.org/power/}} is not only responsible for the SPECpower\_ssj~2008 benchmark, but also for the definitions and tool infrastructures for power measurements~\cite{specpower_andtools}, e.g., the ptdaemon interface, SERT suite, and the Chauffeur Worklet Development Kit, which are also used for other benchmarks and certifications~\cite{energystar3}.
Recently, members of SPEC concluded that the workload of SPECpower\_ssj~2008 does not represent the current demands and described the next version of the benchmark, the still unreleased \specpowernext~\cite{specpower_next,specpower_next2}. %
It will use different technologies for interfaces and measurement handling, but it will also use different workloads targeted at accelerators and CPUs.

\ifx\iflatexml\undefined
\begin{figure}[tp!]
  \centering
  \includegraphics[width=\columnwidth]{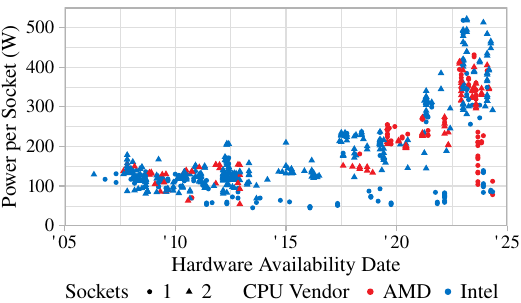}
  \caption{\label{fig:max_power_trend}Power consumption (per socket) at full load trend}
\end{figure}
\else
\fi

As dataset for this paper, we download all \num{\parsedTotalExamined} \texttt{.txt} result files,\footnote{\url{https://www.spec.org/power_ssj2008/results/power_ssj2008.html}} and extract information for hardware and software stack, as well as performance and power measurement results.
We check the consistency, filtering runs that have not been accepted by SPEC~(\num{\parsedRejectedBySPEC}),
runs with ambiguous (\num{\parsedDateAmbiguous}) or implausible~(\num{\parsedDateImplausible}) dates, ambiguous CPU names~(\num{\parsedCPUAmbiguous}), or missing node count~(\num{\parsedNodeCountMissing}), as well as submissions where reported core/thread counts are inconsistent~(\num{\parsedCoreThreadsInconsistent}) or implausible~(\num{\parsedCoreThreadsImplausible}).
This leaves a dataset of~\num{\parsedTotalAccepted} successfully parsed runs.
Each run has four associated dates:
\begin{enumerateinline}
\item The test date,
\item the submission date, as well as
\item hardware and
\item software availability dates.
\end{enumerateinline}
As we discuss trends in hardware, we use the hardware availability date, which indicates the month at which the system became \textcquote{specpower_result_fields}{generally available}.
Hence, even though the earliest results were published in 2007, some runs are associated with earlier dates.

\figref{feature_share_over_years_unfiltered} shows general trends over time.
For the whole duration from \unfilteredYearFirst{} to \unfilteredYearLast{}, an average of \num{\unfilteredRunsPerYearAll} runs were submitted per year.
Between \unfilteredYearDipFirst{} and \unfilteredYearDipLast{}, this drops to \num{\unfilteredRunsPerYearDipMean} runs per year.
The increased number of submissions from \unfilteredYearDipAfterLast{} onward coincides with an increase in submissions using Linux (from \qty{\unfilteredLinuxPercentageUpToDipEnd}{\percent} before \unfilteredYearDipAfterLast{} to \qty{\unfilteredLinuxPercentageAfterDipEnd}{\percent} after \unfilteredYearDipAfterLast{}), and an increase in submissions using AMD processors (from \qty{\unfilteredAMDPercentageUpToDipEnd}{\percent} to \qty{\unfilteredAMDPercentageAfterDipEnd}{\percent}).
The latter observation can be explained by AMD's introduction of its EPYC server CPUs in 2017.

\ifx\iflatexml\undefined
\begin{figure}[t!]
  \centering
  \includegraphics[width=\columnwidth]{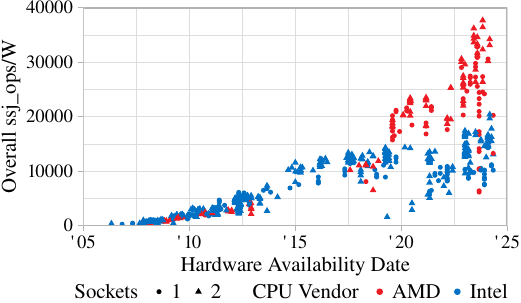}
  \caption{\label{fig:efficiency_trend}Overall efficiency trend}
\end{figure}
\else
\fi

To keep the systems within the dataset comparable, we exclude uncommon configurations:
Runs with CPUs made by neither Intel nor AMD (\num{\unfilteredRunsNeitherAMDNorIntel}), and all runs not on server or workstation CPUs\footnote{I.e. CPUs marketed neither as \emph{Xeon}, \emph{Opteron}, nor \emph{EPYC}.} (\num{\unfilteredRunsNotSever}) are filtered.
Finally, we remove runs with more than one node or more than two sockets (\num{\unfilteredRunsOnlyCommonCfgs}).
After all filtering, \num{\unfilteredRunsAfterFiltering} runs remain as the base for all further analysis.

\section{Performance and Efficiency Trends}
\ifx\iflatexml\undefined
\else
\begin{figure}[tp!]
  \centering
  \includegraphics[width=\columnwidth]{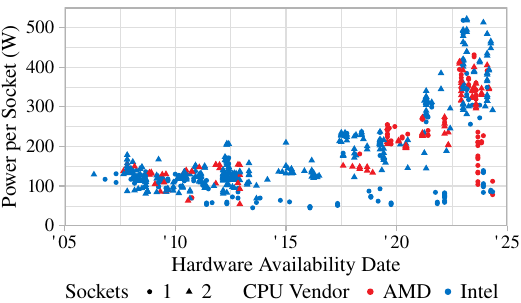}
  \caption{\label{fig:max_power_trend}Power consumption (per socket) at full load trend}
\end{figure}
\fi

Across all runs, the maximum power consumed rapidly increases over the years, as \figref{max_power_trend} depicts.
This trend is consistent across both AMD and Intel, although the spread increased in recent years.
The dataset shows a broad insight into both low and high Thermal Design Power (TDP) processors.
Overall, this upward trend of TDP cannot continue indefinitely, as cooling infrastructure has to be scaled with the power consumed, with air cooling becoming unfeasible at around \qty{400}{\watt}~TDP~\cite{ashrae2021b}.

The increase in power consumption per socket is most pronounced at full load (\qty{100}{\percent}),
where the mean since \powerRecentYearsFirst{} increased \num{\sim \powerFullLoadFactor}x compared to runs up to \powerOldYearsLast{} (\qty{\powerOldYearsMeanFullLoadPower}{\watt} to \qty{\powerRecentYearsMeanFullLoadPower}{\watt}).
However, power consumption at all other load levels increases as well, e.g., by \num{\sim \powerTwentyPercentLoadFactor}x at \qty{20}{\percent} or by \num{\sim \powerSeventyPercentLoadFactor}x at \qty{70}{\percent} load using means across the same time spans.

\ifx\iflatexml\undefined
\else
\begin{figure}[t!]
  \centering
  \includegraphics[width=\columnwidth]{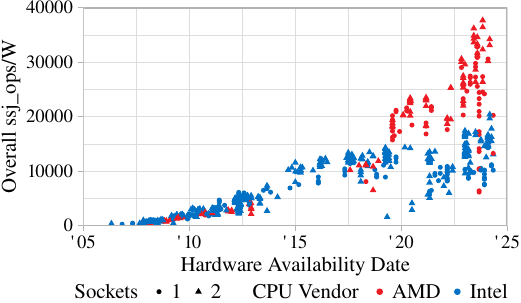}
  \caption{\label{fig:efficiency_trend}Overall efficiency trend}
\end{figure}
\fi

Dividing the achieved performance (\unit{\ssjops} rate) by the mean power consumption results in the energy efficiency in \unit{\ssjops\per\watt}.
Across all runs and load levels, this energy efficiency improved over the years, as \figref{efficiency_trend} shows.\footnote{The overall efficiency (overall \unit{\ssjops\per\watt}) for \specpower is defined as $\sum{\unit{\ssjops}}/\sum{P}$ across all load levels including active idle~\cite{specpower_result_fields}.}
Here, AMD emerges as the driver of the upward energy efficiency trend, in particular from $\sim$2018 onward.
Even though Intel's efficiency is also growing, out of the \num{100} most efficient runs \num{\effRunsByAMDInTopHundred} use AMD processors.
Similarly to the overall score, the runs record the achieved \unit{\ssjops} and average power at every load level, which enables us to compute the efficiency per load level.
We then scale this to the efficiency at full load, yielding the \emph{relative efficiency} per load level.
A relative efficiency \num{<1} is less, \num{>1} more efficient than full load; a relative efficiency of \num{1} at all load levels essentially corresponds to \emph{energy proportionality}.
We summarize these relative efficiencies from \qtyrange{60}{90}{\percent} load, binned by CPU vendor and year in \figref{efficiency_load_levels_violin}.

\begin{figure}[tp!]
  \centering
  \includegraphics[width=0.95\columnwidth]{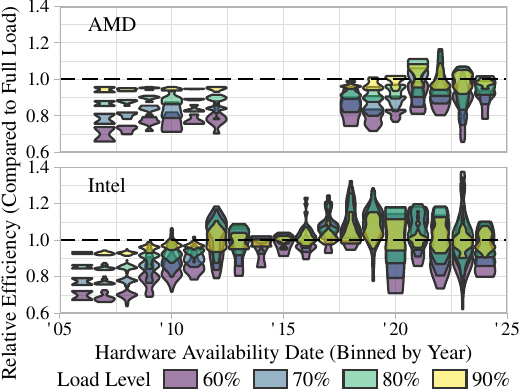}
  \caption{\label{fig:efficiency_load_levels_violin} Distribution of relative efficiency at \qtyrange{60}{90}{\percent}, binned by year and CPU vendor.
  }
\end{figure}

\ifx\iflatexml\undefined
\begin{figure}[bp!]
	\includegraphics[width=\columnwidth]{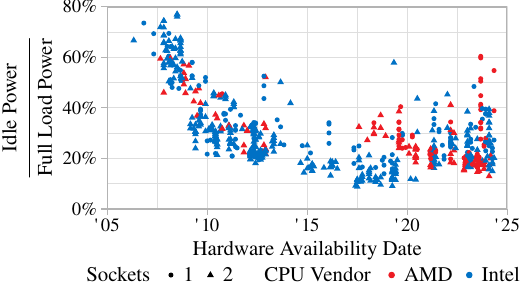}
	\caption{Idle power consumption trend}
	\label{fig:idle_power_fraction_trend}
\end{figure}
\else
\fi

In the early years, lower load is consistently less efficient compared to full load.
Over time, the relative efficiency approaches \num{1} also for lower load levels.
Since 2012 Intel systems have a mean relative efficiency \num{>1} with all load levels \qty{\ge 70}{\percent}, but from 2017 on, we observe a regression back to \num{\sim 1}.
This likely stems from overlapping related effects of increased power reduction at low lower load, but also the use of inefficient turbo states at full load, which were particularly pronounced around 2017.
For AMD, the relative efficiency approaches \num{1} around 2021.
Even though there are still visible differences between AMD and Intel results from 2021 onward, both have a large spread:
No CPU vendor has a universally better relative efficiency at any given load point -- their energy proportionality is diverse.
When comparing absolute instead of relative efficiency however, AMD systems still clearly outperforms Intel systems (cf. \figref{efficiency_trend}).

\section{Idle Power Trend Analysis}
\label{sec:idle_power_trends}

In addition to the full and partial load, the power is measured for an active idle interval.
During active idle, the SUT is ready to perform work, but no transactions are being processed.
Partial load configurations already allow for significant power reduction by leveraging techniques such as Dynamic Voltage and Frequency Scaling (DVFS) and core C-states~\cite{hackenberg_haswell_2015}.
In active idle, the power can be reduced further by powering down shared components, e.g., implemented by package C-states (see~\cite[Chapter~2]{Gough2015}).

The active idle power consumption is particularly relevant for High-Performance Computing (HPC) systems.
HPC systems strive to maximize utilization and may, on average, have a higher utilization than other data center applications. %
However, if no batch job is executed on an HPC node, its load level is truly \qty{0}{\percent}.
In contrast, e.g., a web service during off-hours has a low load level that is typically still \qty{> 0}{\percent}.

\ifx\iflatexml\undefined
\begin{figure}[bp!]
	\includegraphics[width=\columnwidth]{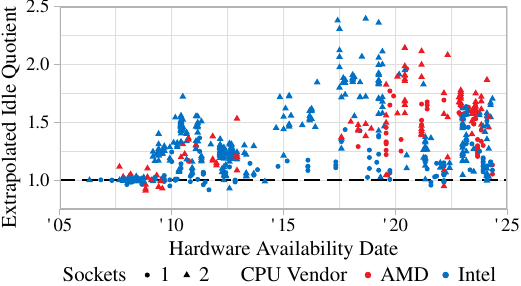}
	\caption{Trend of extrapolated vs measured active idle power}
	\label{fig:extrapolated_idle_trend}
\end{figure}
\else
\fi

The early runs of \specpower in \figref{idle_power_fraction_trend} show the widespread introduction of power-saving mechanisms targeting active idle:
From the earliest runs in \idleFirstYear{}, where idle consumes a mean \qty{\idleFirstYearMeanIdleFractionPercent}{\percent} power compared to full load (the \emph{idle fraction}), the yearly mean drops to its minimum of \qty{\idleMinimumIdleFractionYearlyPercent}{\percent} in \idleMinimumIdleFractionYearlyYear{}.
Since then, the yearly mean idle fraction has increased again to \qty{\idleMaximumIdleFractionYearlyRegressionPercent}{\percent} in \idleMaximumIdleFractionYearlyRegressionYear{}, marking a regression in idle-specific power optimizations.
Intel seems more affected:
In \figref{idle_power_fraction_trend} Intel's runs follow an upward trend, whereas AMD has a slightly falling trend -- although there are both low and high idle fraction systems from both CPU vendors.

\ifx\iflatexml\undefined
\else
\begin{figure}[bp!]
	\includegraphics[width=\columnwidth]{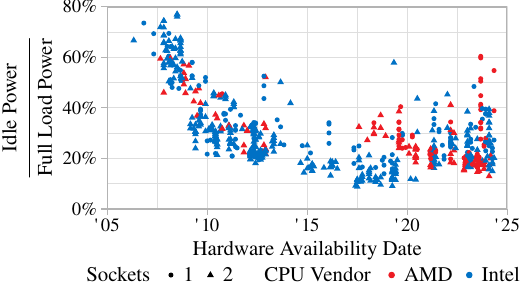}
	\caption{Idle power consumption trend}
	\label{fig:idle_power_fraction_trend}
\end{figure}
\fi

\ifx\iflatexml\undefined
\begin{table*}[tp!]
\centering
\footnotesize
\caption{Comparison of two dual processor Lenovo systems, for the benchmarks SPECpower\_ssj~2008, \speccpu Floating Point Rate Base, and \speccpu Integer Rate Base; \emph{Factor} refers to the relative AMD/Intel performance difference}
\begin{tabular}{| l| l| l| l| l| l| l| l| l |}
\bottomrule
\textbf{Benchmark} & \textbf{Result} & \textbf{Factor} & \textbf{System} & \textbf{CPU} & \textbf{TDP} & \textbf{Date} & \textbf{OS} & \textbf{RAM} \\ 
\hline
power\_ssj~2008 & 15112 & 1 & \multirow{3}{*}{SR650 V3} & \multirow{3}{2.1cm}{Intel Xeon Platinum 8490H 1.90 GHz} & \multirow{3}{*}{ \qty{350}{\watt}} & \multirow{3}{*}{Feb 23} & Windows Server 2019 Datacenter & 256 \\ 
\cline{1-3} \cline{8-9}
CPU 2017 FP & 926 & 1 &  & & & & SUSE Linux Enterprise Server 15 SP4 & 512 \\ 
\cline{1-3} \cline{8-9}
CPU 2017 Int & 902 & 1 &  & & & & Red Hat Enterprise Linux release 9.0 (Plow) & 512 \\ 
\hline
power\_ssj~2008 & 31634 & 2.09 & \multirow{3}{*}{SR645 V3} & \multirow{3}{2.1cm}{AMD EPYC 9754 2.25 GHz} & \multirow{3}{*}{ \qty{360}{\watt}} &  \multirow{3}{*}{Aug 23} & Windows Server 2022 Datacenter & 384 \\ 
\cline{1-3} \cline{8-9}
CPU 2017 FP & 1420 & 1.53 & & & & & SUSE Linux Enterprise Server 15 SP4 & 1536 \\ 
\cline{1-3} \cline{8-9}
CPU 2017 Int & 1830 & 2.03 &  & & & & SUSE Linux Enterprise Server 15 SP4 & 1536 \\ 
\toprule
\end{tabular}
\label{tab:spec_comparison}
\end{table*}
\else
\fi

In an attempt to explain some of this recent development, we explored possible correlations between various run features, including the idle fraction.
This exploration of runs since \lineupsCutoffStartYear{} showed that the CPU vendor lineups, as well as submitted runs affect many features, confounding possible correlations.
Most prominently, the core count of AMD (mean \num{\lineupsMeanCoresAMD}) is greater than that of Intel (mean \num{\lineupsMeanCoresIntel}).
A more subtle example is the nominal frequency, where AMD and Intel share the same mean (\qty{\sim \lineupsFreqGHzMeanAMD}{\giga\hertz}) but differ by spread (standard deviation \qty{\lineupsFreqGHzSDAMD}{\giga\hertz} vs \qty{\lineupsFreqGHzSDIntel}{\giga\hertz}).
Our correlation analysis to explain the recent development of the idle fraction remains inconclusive.

To better understand idle power optimizations, we introduce the \emph{extrapolated} active idle power consumption:
We extrapolate the power consumed at active idle through linear regression from the power consumed at \qty{20}{\percent} and \qty{10}{\percent} load.
The result represents the power consumption during active idle if there would be no specific optimizations for full idle (rather than just individual idle cores).
We then divide this extrapolated by the measured active idle power consumption and plot it over time in \figref{extrapolated_idle_trend}.
We refer to this quotient as \emph{\eiq}.
Higher values correspond to more effective idle-specific power optimization, \num{1} corresponds to none at all.
However, higher values might also indicate a worse energy proportionality at low loads.
Although \figref{extrapolated_idle_trend} has an upward trend overall, there is a large spread, in particular in newer runs.
Idle-specific optimizations are not universally effective in recent runs.

\ifx\iflatexml\undefined
\else
\begin{figure}[bp!]
	\includegraphics[width=\columnwidth]{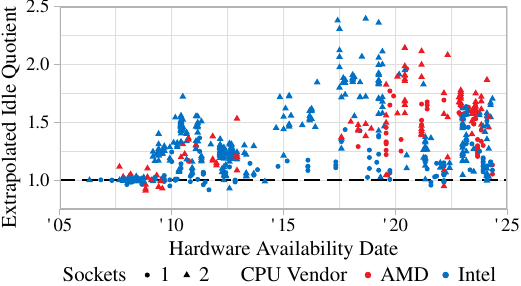}
	\caption{Trend of extrapolated vs measured active idle power}
	\label{fig:extrapolated_idle_trend}
\end{figure}
\fi

The reasons behind this trend are obscured by two indistinguishable mechanisms.
On the one hand, we suspect that processor architectures have an increasingly large share of power being used by shared resources, such as caches and on-chip communication, but also corresponding idle power-saving implementations for these resources.
This effect could increase the \eiq if the latter effectively benefits from the energy-saving techniques.
On the other hand, we speculate that it is becoming more difficult to effectively leverage idle power-saving mechanisms.

Consider, for instance, background tasks that are replicated for each logical CPU.
Their activity prevents the system from fully utilizing idle states for short times each.
With increasing core counts in recent processor generations, more of those tasks are running, reducing the relative time spent in the most efficient idle states.
The first effect -- architecturally low measured active idle -- can explain an increased ceiling of the \eiq.
The second effect -- more difficult effective idle -- can explain the large variation.
From the quantitative data alone, we cannot fully distinguish the compound effects.

\section{Limitations and Generalization}
\label{sec:limitations}
Since we only evaluate data from one benchmark, there is an argument to be made regarding the possibility of generalizing our observed energy efficiency trends.
We evaluated other benchmarks that include power/efficiency:
The TOP500/Green500~\cite{green500} are of limited value with respect to single-node performance due to their additional complexity (e.g., scalability challenges, inter-node networks).
\specomp~\cite{SPEComp2012} and \specaccel~\cite{specaccel} perform floating-point heavy parallel workloads, representing typical parallel scientific workloads.
However, their power measurement support never gained track with submitters; only \num{8} and \num{27} submissions include them, respectively.
Other benchmarks, such as \speccpu~\cite{speccpu2017}, include much more general workloads than \specpower, with equally rich and well-reviewed publicly available datasets, but lack the power measurements required for efficiency analyses.

\ifx\iflatexml\undefined
\else
\begin{table*}[tp!]
\centering
\footnotesize
\caption{Comparison of two dual processor Lenovo systems, for the benchmarks SPECpower\_ssj~2008, \speccpu Floating Point Rate Base, and \speccpu Integer Rate Base; \emph{Factor} refers to the relative AMD/Intel performance difference}
\begin{tabular}{| l| l| l| l| l| l| l| l| l |}
\bottomrule
\textbf{Benchmark} & \textbf{Result} & \textbf{Factor} & \textbf{System} & \textbf{CPU} & \textbf{TDP} & \textbf{Date} & \textbf{OS} & \textbf{RAM} \\ 
\hline
power\_ssj~2008 & 15112 & 1 & \multirow{3}{*}{SR650 V3} & \multirow{3}{2.1cm}{Intel Xeon Platinum 8490H 1.90 GHz} & \multirow{3}{*}{ \qty{350}{\watt}} & \multirow{3}{*}{Feb 23} & Windows Server 2019 Datacenter & 256 \\ 
\cline{1-3} \cline{8-9}
CPU 2017 FP & 926 & 1 &  & & & & SUSE Linux Enterprise Server 15 SP4 & 512 \\ 
\cline{1-3} \cline{8-9}
CPU 2017 Int & 902 & 1 &  & & & & Red Hat Enterprise Linux release 9.0 (Plow) & 512 \\ 
\hline
power\_ssj~2008 & 31634 & 2.09 & \multirow{3}{*}{SR645 V3} & \multirow{3}{2.1cm}{AMD EPYC 9754 2.25 GHz} & \multirow{3}{*}{ \qty{360}{\watt}} &  \multirow{3}{*}{Aug 23} & Windows Server 2022 Datacenter & 384 \\ 
\cline{1-3} \cline{8-9}
CPU 2017 FP & 1420 & 1.53 & & & & & SUSE Linux Enterprise Server 15 SP4 & 1536 \\ 
\cline{1-3} \cline{8-9}
CPU 2017 Int & 1830 & 2.03 &  & & & & SUSE Linux Enterprise Server 15 SP4 & 1536 \\ 
\toprule
\end{tabular}
\label{tab:spec_comparison}
\end{table*}
\fi

To assess the similarity to floating-point workloads, we resorted to screening the \speccpu results for recent runs of similar class CPUs with similar TDP from the same vendor that are also available in our \specpower dataset.
\tabref{spec_comparison} shows one example of two Lenovo nodes with Intel/AMD CPUs, both powered by \qty{1100}{\watt} power supply units, and compares the results of \specpower and \speccpu Rate Base (throughput).
In accordance with our expectations, the relative performance difference between the two systems is similar for \specpower and \speccpu integer, while AMD's performance advantage in the \speccpu floating-point benchmark suite is less pronounced.
The integer-heavy \specpower workload favors AMD CPUs, while Intel's 2x advantage in AVX register width reduces the performance gap for floating-point calculations on wide vectors.
Therefore, the observed energy efficiency trends can not be generalized to floating-point workloads.

Our results can also not be generalized to accelerator-based systems.
While alternative benchmarks can make use of such devices (Green500, \specaccel), we did not consider them for the reasons mentioned previously.

\section{Conclusion and Future Work}
\label{sec:summary}

Our analysis of 16 years of \specpower benchmark results shows continuous increases in \emph{power consumption} of x86 processors.
While there are certainly physical limitations to this growth, they are not yet visible in the data.

\emph{Energy efficiency} also increases continuously and substantially, with AMD clearly providing superior efficiency in recent years.
Due to the integer-heavy properties of the \specpower benchmark, this observation may not be generalized to more floating-point-intensive workloads.
We also observe a positive trend towards better \emph{energy proportionality} for both CPU vendors; although this trend is not universal.

Our analysis of \emph{active idle power} data shows a conclusive trend towards lower consumption between 2006 and 2017, driven by the introduction of successively more effective sleep state mechanisms.
Since then, a substantial share of runs show a regression in idle-specific power optimizations.
We believe that the high variation in the results serves as an indication that particular attention should be paid to practical active idle power in the hardware selection and system operation.
Especially for systems that may spend substantial time in active idle, such as HPC systems, idle power optimizations can improve economical and ecological performance.

The major \emph{limitations} of this analysis are a lack of data for floating-point workloads, for other processor architectures such as ARM, and for accelerators such as AMD or NVIDIA graphics processing units (GPUs). 
The latter will hopefully be addressed by the \specpowernext benchmark~\cite{specpower_next}.
This would be the industry-standard, vendor-driven benchmark, filling an important gap and enabling \emph{future work} on GPU energy efficiency trend analysis.

\section*{Acknowledgments}
This work is supported in part by the German National High Performance Computing (NHR@TUD), funded in equal parts by the state of Saxony and the Federal Ministry of Education and Research.
Additionally, this work is supported by the Federal Ministry of Education and Research via the EECliPs research project (16ME0602).
The authors want to express their gratitude to everyone involved in creating the used \specpower dataset, from the contributors at SPEC creating the benchmark to the countless submitters.
Further, we thank Florian Mros for his work expanding the parsing scripts.

\section*{Availability}
We provide all scripts to reproduce this paper (parsing, analysis, and plotting), together with all data (raw and processed) online~\cite{artifacts}.

\flushcolsend

\bibliographystyle{IEEEtran}

\bibliography{paper}

\end{document}